\documentclass[letterpaper]{article} 
\usepackage{aaai24}  
\usepackage{times}  
\usepackage{helvet}  
\usepackage{courier}  
\usepackage[hyphens]{url}  
\usepackage{graphicx} 
\usepackage{natbib}  
\usepackage{caption} 
\usepackage{enumitem}
\usepackage[ruled,vlined,linesnumbered]{algorithm2e}
\usepackage{amsmath}
\usepackage{amsfonts}
\usepackage{multirow}
\usepackage{tabularx}
\usepackage{makecell}
\DeclareMathOperator*{\argmax}{arg\,max} 
\usepackage{color}

\title{Leveraging Opposite Gender Interaction Ratio as a Path towards \\Fairness in Online Dating Recommendations Based on User Sexual Orientation}
\author {
    Yuying Zhao\textsuperscript{\rm 1},
    Yu Wang\textsuperscript{\rm 1},
    Yi Zhang\textsuperscript{\rm 1},
    Pamela Wisniewski\textsuperscript{\rm 1},
    Charu Aggarwal\textsuperscript{\rm 2},
    Tyler Derr\textsuperscript{\rm 1}
}
\affiliations {
    \textsuperscript{\rm 1}Vanderbilt University\\
    \textsuperscript{\rm 2}IBM T. J. Watson Research Center\\
    \{yuying.zhao, yu.wang.1, yi.zhang, pam.wisniewski, tyler.derr\}@vanderbilt.edu,
    charu@us.ibm.com
}

\begin{document}

\maketitle

\begin{abstract}
Online dating platforms have gained widespread popularity as a means for individuals to seek potential romantic relationships. While recommender systems have been designed to improve the user experience in dating platforms by providing personalized recommendations, increasing concerns about fairness have encouraged the development of fairness-aware recommender systems from various perspectives (e.g., gender and race). However, sexual orientation, which plays a significant role in finding a satisfying relationship, is under-investigated. To fill this crucial gap, we propose a novel metric, Opposite Gender Interaction Ratio (OGIR), as a way to investigate potential unfairness for users with varying preferences towards the opposite gender. We empirically analyze a real online dating dataset and observe existing recommender algorithms could suffer from group unfairness according to OGIR. We further investigate the potential causes for such gaps in recommendation quality, which lead to the challenges of group quantity imbalance and group calibration imbalance. Ultimately, we propose a fair recommender system based on re-weighting and re-ranking strategies to respectively mitigate these associated imbalance challenges. Experimental results demonstrate both strategies improve fairness while their combination achieves the best performance towards maintaining model utility while improving fairness.
\end{abstract}

\section{Introduction}\label{sec-intro}
Online dating has grown increasingly popular and is now a leading way of finding romantic partners and even meeting new friends~\cite{rosenfeld2019disintermediating}. 
For example, in 2022 it was estimated that 30\% of U.S. adults had used online dating and even upwards of 51\% among lesbian, gay or bisexual adults~\cite{mcclain2023looking}. To accommodate this growing demand, various platforms have emerged, e.g., OkCupid, Tinder, and Grindr. With the booming of users, the challenge of information/choice overload~\cite{pronk2020rejection} and unawareness~\cite{finkel2012online} have made recommender systems (RS) even more important, which learn user preferences via their behaviors on the platform. This ultimately provides users with recommended partners that hopefully match their interests and significantly enhance their experience~\cite{xia2015reciprocal}.

However, while RS improve user satisfaction, fairness concerns still exist if systems are solely designed to maximize overall utility. For example, race-related fairness has been investigated to decrease racial homogamy via agent-based model interventions on online dating platforms~\cite{ionescu2021agent}. Additionally, in online dating, different gender identities have diverse characteristics, motivations, preferences, etc~\cite{abramova2016gender}. Thus, if ignored, this generally leads to an inherent distinction in recommendation quality across gender identities, which has motivated past work on gender-aware system modifications to ensure equitable outcomes~\cite{zheng2018fairness}. Nevertheless, although the aforementioned fairness perspectives are crucial and provide additional consideration beyond utility, another important sensitive user characteristic associated with dating is their sexual orientation, but less commonly discussed in the literature. 

In one of the most basic forms, the satisfaction of a recommendation is contingent upon users' sexual orientations and the gender identity of those being recommended to them. Various sexual orientations indicate users' sexual preferences, including but not limited to homosexual individuals who prefer the same gender as their romantic partner, heterosexual individuals who prefer the opposite gender, and bisexual individuals who are attracted to both genders. However, even for bisexual individuals the spectrum as to their preference on dating certain genders varies, raising further challenges in the recommendation system. To exacerbate this issue, studies have shown that personal experiences with online dating significantly differ by sexual orientation~\cite{rosenfeld2010meeting,finkel2012online}. 

\textit{With diverse preferences and demands, could users with various sexual orientations be treated similarly?} Unfortunately, unfairness would be likely to exist for the heteronormativity assumption. Specifically, heterosexual users are generally the majority of dating applications (if without specific design, such as Grindr, which is designed specifically for the LGBTQ community), and RS inherently tend to perform better for users aligned with the preferences/behaviors of the majority while compromising the performance of the minority; thus, leading to the unfairness. However, while these minority groups by definition are lower in percentage, they are also increasing in size~\cite{jones2021lgbt} and nearly twice as likely to report using an online dating platform~\cite{mcclain2023looking}. This indicates despite being a smaller proportion, minority groups constitute a substantial number of users who might have a higher desire for online dating services and deserve high-quality recommendations.

Although the above discussion strengthens the motivation and need to investigate the potential unfairness of RS in online dating according to users' sexual orientations, it is nontrivial to study this problem due to the following challenges: (C1) The lack of knowledge of accurate sexual orientation. While platforms could allow users to specify their sexual orientation, some users might be reluctant to specify their sexual orientation due to privacy concerns or lack of suitable selection options on the dating platform; sexual orientation alone is insufficient for a high-quality recommendation, especially for bisexual users (e.g., if a bisexual user tends to prefer mostly the opposite gender, but the system recommends primarily users of the same gender, it would result in unsatisfying recommendation performance); sexual fluidity is prevalent, and users' sexual orientation might change over time. (C2) Improving fairness without compromising overall utility is a long-standing issue in fairness-related studies and has no established answers till now~\cite{li2022fairness}.

To address these challenges, this work presents the initial endeavor to investigate fairness of online RS from sexual orientation perspective. To obtain knowledge about sexual orientation, rather than directly classifying users into various categories which are unreliable due to a lack of user profiles in our dataset, we extract an interaction-based metric called Opposite Gender Interaction Ratio (OGIR), which serves as an implicit indicator (i.e., if an individual interacts with both genders, but mostly with the opposite gender, they are likely bisexual but with a stronger preference to the opposite gender). After obtaining OGIR, we divide users into groups where groups have different levels of OGIR, indicating their diverse preferences towards the opposite gender. Given groups, we empirically investigate and verify the existence of group unfairness in existing RS where groups are treated differently in terms of recommendation quality. To mitigate the performance gap among groups, we identify two potential causes: group quantity imbalance and calibration imbalance~\cite{steck2018calibrated}. Correspondingly, we propose an in-processing re-weighting strategy and a post-processing re-ranking strategy. Experimental results show that both strategies improve fairness and have their unique advantages. When utilized together, these strategies lead to best performance in improving fairness while maintaining utility performance. Our main contributions are:
\begin{itemize}[leftmargin=*]
    \item 
    We observe the presence of consistent group unfairness based on Opposite Gender Interaction Ratio (OGIR), which is related to users' sexual orientation, in multiple recommenders in a real-world online dating dataset;
    \item 
    We identify two potential causes for group unfairness: group quantity and calibration imbalance. Correspondingly, we design re-weighting and re-ranking strategies;
    \item Experiments show that both strategies are effective at reducing the recommendation quality gap across groups divided by OGIR. 
    Furthermore, combining the two strategies results in the best performance. 
\end{itemize}

\section{Related Work}
\label{sec-related_work}
\subsection{Recommender Systems in Online Dating}
RS serves as an effective solution to tackle information overload by delivering personalized recommendations. There have been numerous works in designing online dating RS, including interaction-based and content-based methods. Most interaction-based methods employ collaborative filtering~\cite{brozovsky2007recommender,krzywicki2010interaction}, which generate recommendations according to user similarities. For instance, collaborative filtering methods had been previously used to estimate the attractiveness rating of user pairs according to the ratings of similar users~\cite{brozovsky2007recommender}. On the other hand, content-based methods utilize user profiles and features for recommendations~\cite{hitsch2010matching,zheng2022you}. For example, Latent Dirichlet Allocation (LDA) has been previously used to learn user preferences~\cite{tu2014online}. Additionally, to satisfy user requirements from both ends, reciprocal recommendation methods are proposed~\cite{pizzato2010recon,xia2015reciprocal}. In summary, these approaches effectively capture user preferences and enhance user experience. Nonetheless, few of them take fairness into account during algorithm development.

\subsection{Fairness in Online Dating}
Although fairness has been extensively studied~\cite{zhao2023fairness,wang2022survey, wang2022improving}, fairness works in online dating are still relatively few. The most related stream of work focuses on promoting fairness among groups of users according to their associated sensitive attribute, with race~\cite{sapiezynski2019quantifying,paraschakis2020matchmaking}, gender~\cite{zheng2018fairness,melchiorre2021investigating}, and religion~\cite{paraschakis2020matchmaking} being among the most commonly studied. For example, a group fairness metric that not only depends on the ranking results but also on the distribution of user attention was proposed to improve racial fairness~\cite{sapiezynski2019quantifying}. In addition, individual fairness metrics have also been developed, such as calibration-based methods to encourage recommending potential partners that match user preferences focusing on race and religion~\cite{paraschakis2020matchmaking}, which shares a similar objective to our research in terms of promoting fairness through calibration, but they focus on conformity to user preferences, while our aim is to mitigate the performance disparity among user groups according to their sexual orientations. Specifically, we also aim to ensure fairness among groups divided based on sensitive attributes, but to the best of our knowledge, this work presents the first endeavor to study fairness from the perspective of sexual orientation and draw connections to imbalanced learning. 
\section{Online Dating Dataset Analysis}
\label{sec-analysis}

\begin{figure*}
    \centering
    \includegraphics[width=0.88\textwidth]{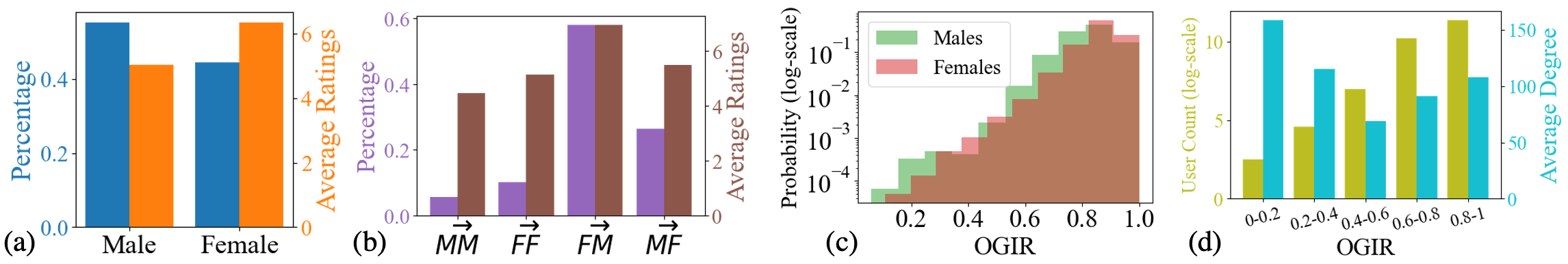}
    \vspace{-0.75ex}
    \caption{Dataset analysis (a) gender identity distribution and their average ratings; (b) interaction type distribution and their average ratings; (c) OGIR distribution of female/male users;  (d) user counts and average degrees according to OGIR.
    }
    \label{fig:distribution}
    \vspace{-1.75ex}
\end{figure*}

In this work, we use a real-world dataset from Líbímseti.cz (which is hosted in the Czech Republic) and is publicly available~\cite{brozovsky2007recommender,kunegis2012online}\footnote{\label{dataset} Dataset links are available at \url{https://github.com/YuyingZhao/Fair-Online-Dating-Recommendation}.}. Unfortunately, many works are unable to make their data public~\cite{zhao2013user,al2017user,xia2015reciprocal}, and other available dating datasets pose limitations. For example, OkCupid and Lovoo\footref{dataset} provide user profiles without interactions. The Speed Dating dataset\footref{dataset} was gathered from experimental speed dating events, but smaller scale and not related to online dating. Therefore, this dataset is particularly valuable as it not only contains user interactions, but also the self-identified gender information of the users, and the platform was not exclusively designed for heterosexual users, which enables the investigation presented in this work.

This section presents a detailed analysis of the Líbímseti.cz dataset, providing additional context for interpreting our empirical results. Overall the dataset~\cite{kunegis2012online} contains $220,970$ users and $17,359,346$ interactions in the form of $(u, v, r)$ tuples where user $u$ rates user $v$ with score $r$ according to $u$'s preference. Some users have filled in their (binary\footnote{This work focuses on binary case, attributed to limited dataset and does not reflect authors' opinions on gender identity.}) gender information, while others' remain unknown. In this study, we concentrate on users who provide gender identity information. The detailed binary gender identity distribution and their corresponding average ratings to other users are shown in Fig.~\ref{fig:distribution}(a). Among the users with gender information, we further explore the types of interactions where one user rates the other, leading to four types [`Male$\rightarrow$Male', `Female$\rightarrow$Female', `Female$\rightarrow$Male', `Male$\rightarrow$Female'] abbreviated as  [$\overrightarrow{\text{MM}}$, $\overrightarrow{\text{FF}}$, $\overrightarrow{\text{FM}}$, $\overrightarrow{\text{MF}}$]. The interaction type distribution and their average ratings are shown in Fig.~\ref{fig:distribution}(b). Based on users' interaction, we count the proportion of each user interacting with opposite genders, measured by opposite gender interaction ratio (OGIR).

\textbf{Opposite Gender Interaction Ratio ($\text{OGIR}$)} for a user defines the ratio of opposite genders among this user's interaction history, which captures the tendency of a user being sexually attracted by users of the opposite gender. Suppose user $u$ has rated $N_u$ users among which $\hat{N_u}$ is the number of individuals from opposite gender with user $u$. Formally, it is defined as: $\text{OGIR}_u = \hat{N_u}/N_u$. By definition, $\text{OGIR}$ lies in the range [0, 1]. Users with $\text{OGIR}$ closer to 0/1 are more toward homosexual/heterosexual.

The histogram of users' OGIR in Fig.~\ref{fig:distribution}(c) shows that most users, regardless of gender, prefer to interact with users of opposite genders. Fig.~\ref{fig:distribution}(b) shows that females ($\overrightarrow{\text{FF}}$ and $\overrightarrow{\text{FM}}$) on average tend to rate higher than males ($\overrightarrow{\text{MM}}$ and $\overrightarrow{\text{MF}}$). Additionally, hetero-interactions (i.e., interaction between different genders, $\overrightarrow{\text{FM}}$ and $\overrightarrow{\text{MF}}$) tend to have higher ratings than homo-interactions (i.e., interaction between the same gender, $\overrightarrow{\text{FF}}$ and $\overrightarrow{\text{MM}}$). 
We also plot the user number and average degree according to OGIR in Fig.~\ref{fig:distribution}(d). The user count aligns with the conclusion from Fig.~\ref{fig:distribution}(c) where majority prefer opposite gender. The degree indicates users with low/high OGIR tend to have more interactions on average. 

To summarize, we draw the following observations: 
\begin{itemize}[leftmargin=*]
    \item Males take up a larger proportion than females, but females tend to rate more frequently than males, leading to a larger proportion of $\overrightarrow{\text{FF}}$ and $\overrightarrow{\text{FM}}$ than $\overrightarrow{\text{MM}}$ and $\overrightarrow{\text{MF}}$. 
    \item Most interactions are between different genders (i.e., $\overrightarrow{\text{FM}}$, $\overrightarrow{\text{MF}}$) while those within same gender also exist (i.e., $\overrightarrow{\text{FF}}$, $\overrightarrow{\text{MM}}$), which indicates the interactions are 
    multi-faceted and (on average) users with OGIR 0 to 0.4 have the highest level of engagement/degree.
    \item Users tend to prefer/ignore
    the opposite gender at varied levels, which indicates that user sexual preferences toward the opposite gender are complex and diverse.
\end{itemize}

\section{Fairness Concerns in Online Dating Recommendations} 
\label{sec-concerns}
In the last section, we analyzed complex user behaviors in a real-world online dating site with an emphasis on the users' opposite gender interaction ratio (OGIR), which provides insight into user sexual orientations according to their historical interactions. In this section, \textit{we seek to study whether users grouped by OGIR, who have diverse levels of preferences toward the opposite gender, would be treated fairly} if a recommender system was to be applied to improve their user experience. Specifically, we first formally define the group unfairness based on the average performance gap between groups, then we perform an initial empirical evaluation on off-the-shelf recommendation algorithms to simulate whether unfairness was to exist if such a recommender system if deployed in the real world. 

\subsection{User-based Group Unfairness}
Following existing literature that fairness can be interpreted as the equality of utility across entities in different groups~\cite{fu2020fairness,li2021user}, we define user-based group unfairness as the difference of recommendation performance across users with different levels of OGIR. Intuitively, a larger gap indicates higher discrimination/lower fairness. In the following, we define how to divide groups based on OGIR and the corresponding unfairness metrics.

\begin{figure*}[tbp]
    \centering
    \includegraphics[width=0.95\textwidth]{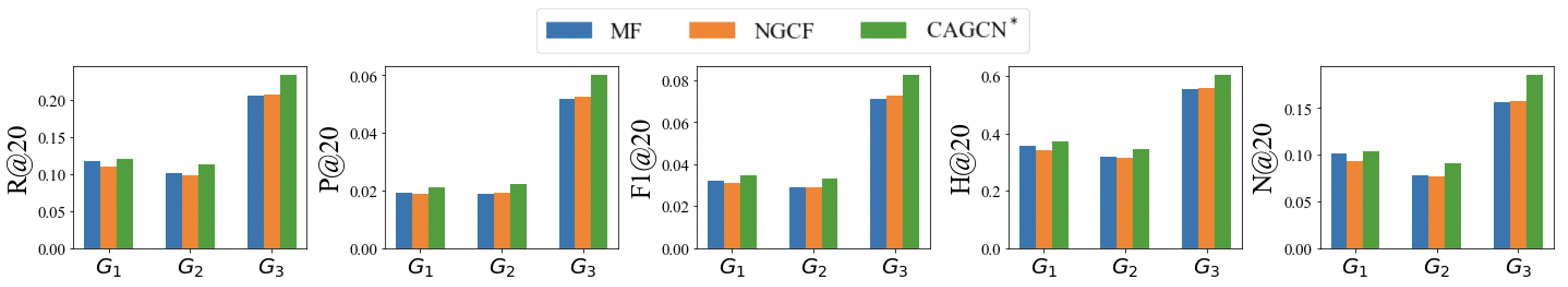}
    \vskip -1ex
    \caption{Utility performance of three models on five metrics, where groups are divided based on even width bins for discretizing OGIR into three groups ($G_1=\{u|\text{OGIR}_u\in [0, \frac{1}{3})\}$ with $G2$, $G3$ similarly defined). $G_3$ consistently has better performance.}
    \label{fig:baselines}
\end{figure*}

\subsubsection{Group Partition}
To quantify such unfairness, we divide users into multiple groups based on their OGIR. Specifically, users in each group are within the same interval of $\text{OGIR}$, and groups have equal interval ranges. As these groups have different levels of OGIR, they separate users based on their diverse preferences toward the opposite gender but do not designate homosexual/bisexual/heterosexual user groups. For this study, we construct a $3$-group partition where groups are denoted as $G_1$, $G_2$, and $G_3$, and have users with $\text{OGIR}$ in ranges $[0, \frac{1}{3}), [\frac{1}{3}, \frac{2}{3})$, and $[\frac{2}{3}, 1]$, respectively.

\subsubsection{User-based Group Unfairness Metric} Our proposed metric measures the discrepancy of recommendation performance among groups $\mathcal{G}$, which is defined as the average performance gap of certain metrics $\text{X}$
(e.g., recall, F1, etc) among group pairs:
\begin{equation}
    \Delta_{\text{X}}(\mathcal{G}) = \frac{1}{Q_X^{ave}}\mathbb{E}_{(G_1, G_2)\in \mathcal{G} \times \mathcal{G}}|Q_X(G_1)-Q_X(G_2)|,
    \label{eq.unfairness}
\end{equation}
where $(G_1, G_2)$ is a unique group pair (i.e., $G_1 \neq G_2$), and $Q_X(G_i)=(\sum_{u \in G_i}q_x(u))/|G_i|$ is the average recommendation performance measured by metric $\text{X}$ of users in the group $G_i$ with $q_x(u)$ being user $u$'s performance according to metric X. The denominator normalizes by the average performance to mitigate the impact of performance scale across metrics where $Q^{\text{ave}}_X = (\sum_{G \in \mathcal{G}}Q_X(G))/|\mathcal{G}|$.

\subsection{Initial Fairness Evaluation}
We evaluate various models to investigate group unfairness. The detailed pre-processing steps and baseline descriptions are in the Appendix. From the evaluation, we observe consistent unfairness and discuss potential ``fixes" which could not work and urge the need for a fair model.
\subsubsection{Evaluation Metrics and Models} 
We include various utility metrics and their corresponding fairness metrics for a comprehensive comparison, including Recall (R@20), Precision (P@20), F1@20, Hit Ratio (H@20), and Normalized Discounted Cumulative Gain (N@20) and their corresponding fairness metrics according to Eq.~\ref{eq.unfairness} ($\Delta_\text{R}@20$, $\Delta_\text{P}@20$, $\Delta_\text{F1}@20$, $\Delta_\text{H}@20$, and $\Delta_\text{N}@20$). For utility/fairness metrics, the higher/lower the value, the better the performance. We evaluate across three representative recommenders, including seminal works and current state-of-the-art: MF~\cite{rendle2012bpr}, NGCF~\cite{wang2019neural}, and CAGCN${\bf ^*}$~\cite{wang2022cagcn}. They are optimized with Bayesian Personalized Ranking (BPR) loss, $\mathcal{L}_{\text{BPR}}$~\cite{rendle2012bpr}. 

\subsubsection{Evaluation Results}
To mitigate the randomness impact for a better comparison, we run the evaluated models $5$ times with different seeds and report the average results. Without specification, the group number is set to $3$. The model selection is based on the average utility score on validation. The average utility result in Fig.~\ref{fig:baselines} shows that generally, $G_3$ has better performance than $G_1$ and $G_2$, indicating that $G_3$ enjoys better recommendation quality. The performance gap among groups is quantified by the proposed unfairness measurement where these models have more than $0.5$ unfairness scores, presenting a consistent unfairness that appears to be algorithm/model-agnostic according to our results.

\subsubsection{Potential Na\"ive ``Fix'' Towards Fairness}
One potential approach to addressing the unfairness issue could be a fairness-aware model selection. For example, one could use $\text{score} = \text{Avg Utility}-\text{Avg Fairness}$. The experiment shows no significant fairness improvement compared with baseline models. This indicates that simply considering fairness in model selection is insufficient for a fair model. Another potential solution would be to train the recommender separately for different groups. However, it presents two challenges. First, it will further exacerbate the data sparsity issue, which would be more severe for the minority than the majority. Secondly, in real world, a user in one group might be interested in a user from another group. Separate training would result in restricted recommendations and a suboptimal outcome. Therefore, both potential na\"ive ``fixes" cannot solve the problem. This raises the requirement of designing a new fair model, which we present in the next section.

\section{Fair Recommender System}
\label{sec-model}
In this section, we analyze potential unfairness from group quantity and calibration imbalance. To mitigate them, we introduce re-weighting and re-ranking strategies.

\subsection{Mitigating Group Quantity Imbalance: \hspace{20ex} Re-weighting Towards Improved Fairness}

\begin{figure}[tbp]
    \centering
    \includegraphics[width=0.46\textwidth]{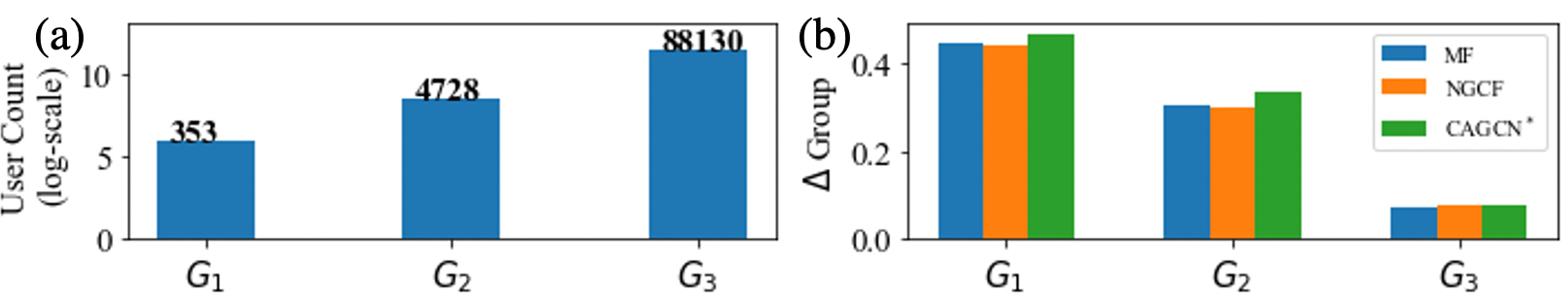}
    \caption{Two potential causes of unfairness (a) group quantity imbalance; (b) group calibration imbalance.}
    \label{fig:unfairness_reasons}
    \vspace{-3ex}
\end{figure}

The issue of class imbalance, where the number/quantity of training instances per class is imbalanced, has been widely investigated across various domains~\cite{johnson2019survey, chawla2002smote}. During the training process, to achieve an overall higher utility performance, the majority class is typically optimized more than the minority class, leading to a performance gap. As shown in Fig.~\ref{fig:unfairness_reasons}(a), the numbers of users in different groups are imbalanced in our setting (i.e., $G_1$ is the majority, $G_2$ and $G_3$ are the minorities). As a consequence, there are performance gaps among majority and minority groups, resulting in unfairness. To mitigate this unfairness, we employ the re-weighting strategy, which has been utilized to address the class imbalance issue. This approach adjusts the focus of training by updating the weights based on the number of users in each group effectively balancing the original loss function accordingly such that equitable emphasis is put on each group when updating the model's parameters.

In traditional $\mathcal{L}_{\text{BPR}}$, each tuple is trained equally without consideration of group size. Generally, as one group (e.g., majority) appears more in the training data during the optimization, the users belonging to this group will achieve better performance as they share common (group-level) user behaviors. To remedy this, we add a weight term for adjustment. The updated loss is as follows:
\begin{equation}
    \mathcal{L}_{\text{BPR}}^{\text{re-weighting}} = - \hspace{-2ex}\sum_{(u, i, j) \in \mathcal{D}} w_{G(u)} \log\sigma(\hat{y}_{ui}-\hat{y}_{uj})+\lambda_{\Theta}\|\Theta\|^2,
    \nonumber
\end{equation}

\noindent with training data $\mathcal{D}=\{(u,i,j)|u\in\mathcal{U}, i\in \mathcal{I}_u^+, j\in \mathcal{I}_u^-\}$, total user set $\mathcal{U}$, user sets $u$ did (not) interacted with $\mathcal{I}_u^+$ ($\mathcal{I}_u^-$),
predicted preference score $\hat{y}_{ui}$, and user $u$'s weight based on $u$'s group, $w_{G(u)}$. Generally, when $u$ belongs to a group with a larger user number (e.g., $G_3$), the weight will be lower than the case when $u$ belongs to one with a smaller user number (e.g., $G_1$ and $G_2$) to promote the training for the minority. Specifically, we utilize $w_{G, p}= \frac{1}{{N_G}^p}$ where $N_G$ is the number of users in the group $G$ and $p$ is for different weight assignments. Compared with the original utility objective, the updated objective considers utility and fairness simultaneously with $p$ balancing two goals. 

\subsection{Mitigating Group Calibration Imbalance: \hspace{20ex} Re-ranking Towards Improved Fairness}

The notion of calibration in recommendation refers to the property that the genre distribution (e.g., Sci-Fi, Romance, etc. in movie recommendation) in the recommendation list should match the distribution in the history interactions~\cite{steck2018calibrated}. A higher-quality calibration means a lower level of inconsistency between the distributions, which indicates that the model can better preserve users' preferences. In dating recommendation, a good calibration requires the ratios of males/females in training and recommendation to be similar. We quantify the calibration score, $\Delta_{\text{User}}(u, \mathcal{R}_u), $ of a user $u$ as the inconsistency between the ratio of female users that are interacted in the training dataset (i.e., $T^F(u)$) and the ratio of females that are recommended in the recommendation list $\mathcal{R}_u$ (i.e., $R^F(\mathcal{R}_u)$). Formally, this is defined with the absolute difference as follows: 
\begin{align*}
    \Delta_{\text{User}}(u, \mathcal{R}_u) = |T^F(u)-R^F(\mathcal{R}_u)|. 
\end{align*}
Then, we quantify the calibration of a group by averaging the calibration scores of the users in that group as follows:
\begin{align*}
    \Delta_{\text{Group}}(G, \mathcal{R}) = \sum\nolimits_{u \in G} \Delta_{\text{User}}(u, \mathcal{R}_u).
\end{align*}

In Fig.~\ref{fig:unfairness_reasons}(b), we calculate the group calibrations (where lower is better) across the baseline models and observe all exhibit group calibration imbalance. It shows an opposite trend with the performance in Fig.~\ref{fig:baselines} that $G_3$ has the lowest calibration score and the highest performance. We posit that utility performance is negatively correlated with calibration scores. Since the trained model is more towards the majority, the ability to preserve the users' preferences is compromised for the minority. Based on this hypothesis, we aim to mitigate the calibration imbalance issue by reducing the inconsistency between the gender ratio of training interactions and the recommendation list by re-ranking strategy. The minority has poor calibration, but this indicates a large space for improvement. Therefore, by ensuring better calibration, it can potentially improve the utility performance of all groups with a larger improvement for the minority, which will lead to a decrease of utility gap and thus improve fairness.

\setlength{\textfloatsep}{2pt}
\SetAlgoNoEnd
\begin{algorithm}[tb]
\footnotesize
 \DontPrintSemicolon
 \KwIn{Recommendation number $K$; user id $u$; trade-off parameter $\lambda$, $u's$ top $K^\prime$ baseline recommendations as candidates $\mathcal{C}_u$} 

 $\mathcal{R}_u = \{\}$
 
 \While{$|\mathcal{R}_u|\leq \text{K} $}{
   $i^* = \text{argmax}_{i \in \mathcal{C}\setminus\mathcal{R}_u} (1-\lambda) S({\mathcal{R}_u \cup \{i\}})- \lambda \Delta_{\text{User}}(u, {\mathcal{R}_u \cup \{i\}})$

    $\mathcal{C}_u={\mathcal{C}_u \setminus \{i^*\}}$
    
   $\mathcal{R}_u = {\mathcal{R}_u \cup \{i^*\}}$
    
}
\KwRet{User $u$'s re-ranked recommendation list $\mathcal{R}_u$}

\caption{Greedy Algorithm for Re-ranking to Mitigate Calibration Imbalance}
\label{alg:greedy}
\end{algorithm}

The re-ranking strategy is a post-processing mechanism to find new recommendations based on the original recommendations from baseline models.
With utility and calibration consideration, we use Maximum Marginal Relevance (MMR) ~\cite{carbonell1998use, steck2018calibrated, zhao2021rabbit} to determine the recommendation list $\mathcal{R}_u^*$ for user $u$, so our objective is formalized as follows:
\begin{equation}
    {\mathcal{R}_u}^* = \argmax_{\mathcal{R}_u, |\mathcal{R}_u|=\text{K}}(1-\lambda)S(\mathcal{R}_u)-\lambda \Delta_{\text{User}}(u, \mathcal{R}_u)
    \label{eq:mmr}
\end{equation}

\noindent The objective is composed of two terms with trade-off parameter
$\lambda\in [0,1]$
(1) the predicted relevance score $\hat{y}_{ui}$ from baseline models related to the utility performance, where $S(\mathcal{R}_u)=\sum_{i\in \mathcal{R}_u}\hat{y}_{ui}$; and (2) the calibration score $\Delta_{\text{User}}(u, \mathcal{R}_u)$. Additionally, as $\Delta_{\text{User}}(u, \mathcal{R}_u) \in [0, 1]$, we rescale the relevance scores so that they fall in the same range.
Solving Eq.~\ref{eq:mmr} NP-hard ~\cite{steck2018calibrated}. We adopt a greedy algorithm~\cite{nemhauser1978analysis} in Algorithm~\ref{alg:greedy}, which finds the approximate solution with $(1-\frac{1}{e})$ optimality guarantee where $e$ is the natural logarithm. To recommend potential partners for a user $u$, Algorithm~\ref{alg:greedy} starts with an empty list with top $K^\prime$ individuals recommended from the original baseline models as the candidate set $\mathcal{C}_u$ and then iteratively adds the optimal individual that obtains the largest score. The algorithm ends when the list reaches length $K$.

\begin{table*}[t!]
\small
\setlength\tabcolsep{2.2pt}
 \caption{Performance comparison of baseline model versus re-weighted model (model$_\text{rw}$). The $\uparrow$ represents the larger the better and $\downarrow$ represents the opposite. The proportion (+/- $\%$) shows the performance improvement/degradation to the baseline model.}
\begin{tabular}{|c|c|c|c|c|c|c|c|c|c|c|c|c|}
\hline
\multirow{2}{*}{Method} & \multicolumn{6}{c|}{Utility Metrics $\uparrow$} & \multicolumn{6}{c|}{Fairness Metrics $\downarrow$} \\\cline{2-13}
  & R@20 & P@20 & F1@20 & H@20 & N@20 & Avg Utility & $\Delta_{\text{R}}@20$ & $\Delta_{\text{P}}@20$ & $\Delta_{\text{F1}}@20$ & $\Delta_{\text{H}}@20$ & $\Delta_{\text{N}}@20$ & Avg Fairness \\

 \hline

MF & 0.2002 & 0.0499 & 0.0690 & 0.5406 & 0.1517 & 0.2023 & 0.4964 & 0.7361 & 0.6397 & 0.3861 & 0.4664 & 0.5449\\
\hline
NGCF & 0.2019 & 0.0508 & 0.0701 & 0.5457 & 0.1527 & 0.2043 & 0.5294 & 0.7577 & 0.6611 & 0.4016 & 0.4961 & 0.5692 \\
\hline
CAGCN$^*$ & 0.2267 & 0.0580 & 0.0798 & 0.5890 & 0.1802 & 0.2267 & 0.5196 & 0.7534 & 0.6562 & 0.3929 & 0.4955 & 0.5635 \\
\hline

MF$_\text{rw}$ & 0.2003 & 0.0503 & 0.0694 & 0.5421 & 0.1472 & 0.2019 (-0.20\%) & 0.3945 & 0.6491 & 0.5447 & 0.3106 & 0.3264 & 0.4450 (+18.33\%) \\
\hline
NGCF$_\text{rw}$ & 0.1932 & 0.0482 & 0.0668 & 0.5287 & 0.1430 & 0.1960 (-4.06\%) & 0.3832 & 0.6274 & 0.5290 & 0.2924 & 0.3187 & 0.4301 (+24.44\%)\\
\hline
CAGCN$^*_\text{rw}$ & 0.2242 & 0.0566 & 0.0781 & 0.5854 & 0.1780 & 0.2244 (-1.01\%) & 0.4928 & 0.7222 & 0.6310 & 0.3718 & 0.4577 & 0.5351 (+5.04\%) \\
\hline

 \end{tabular}
\label{tab-reweight-result}
\end{table*}

\section{Experiments}
In this section, we conduct experiments to verify the effectiveness of Re-weighting and Re-ranking strategies\footnote{Source code is available at:  \url{https://github.com/YuyingZhao/Fair-Online-Dating-Recommendation}} under the setting of $K=20$ and $K^\prime=100$. We aim to answer two main research questions for both strategies.

\begin{itemize}[leftmargin=*]
    \item \textbf{RQ1}: How well can proposed strategies improve fairness while not significantly decreasing utility performance?
    \item \textbf{RQ2}: What are the impacts of the hyperparameters?
\end{itemize}
To answer these questions, we first report the re-weighting and re-ranking results. We also report the result of applying them jointly. After analyzing the results, we present a discussion about these strategies in the end.

\subsection{Experimental Results with Re-weighting}
\label{sec.reweight}

\subsubsection{Re-weighting Performance}
\begin{figure}[t]
    \centering
    \includegraphics[width=0.42\textwidth]{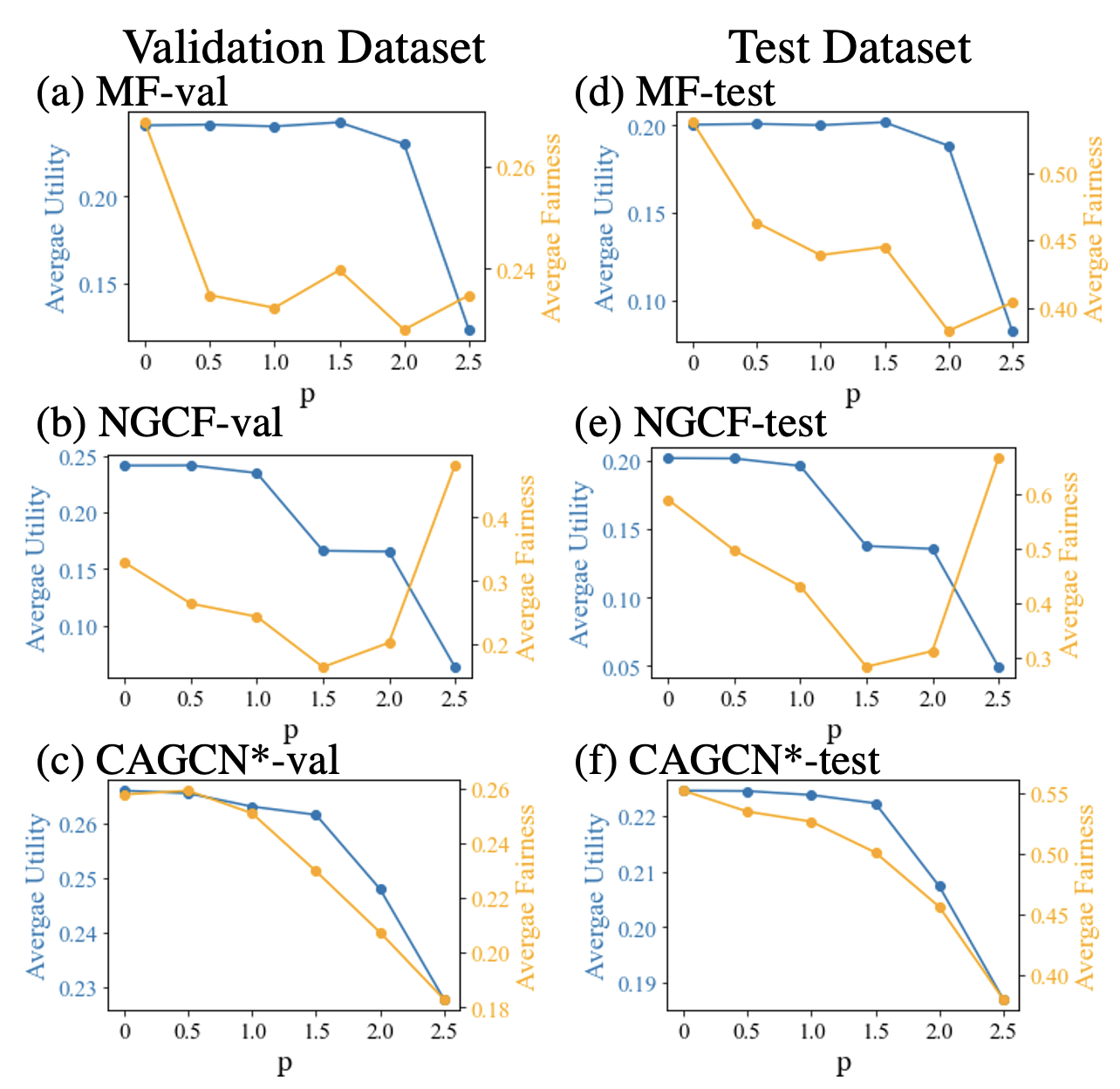}
    \vspace{-2.5ex}
    \caption{Analysis on the utility and fairness performance impacts associated with the re-weighting hyperparameter $p$.}
    \label{fig:test_reweight}
    \vspace{1ex}
\end{figure}

Table~\ref{tab-reweight-result} shows the test performance of specific $p$ with standard deviations omitted (always less than $0.02$). We select $p$ based on the validation dataset, where we plot the validation curve as shown in Fig.~\ref{fig:test_reweight}(d-f) and select $p$ before the sharp decrease in utility performance to avoid a large compromise in the overall performance (i.e., $1.5$ for MF, $1.0 $ for NGCF, and $0.5$ for $\text{CAGCN}^*$). Other strategies can be applied to select the best hyperparameter based on the validation curve, where the tradeoff between fairness and utility can be clearly observed. Thus, platforms can pick the hyperparameter based on their demands. In this way, the model selection is more flexible. Compared with the sensitivity analysis in the following subsection, we would find that the validation curve generally matches the trend of the test curve, which validates that it is reliable to select the best hyperparameter based on the validation record. From Table~\ref{tab-reweight-result}, we observe that with the re-weighting strategy, for each method, the fairness improves with a little sacrifice of utility performance. NGCF has the best improvement in fairness (i.e., $24.44\%$), while CAGCN$^*$ has the smallest improvement (i.e., $5.04\%$). 

\subsubsection{Sensitivity Analysis of $p$}
\label{sec-ablation-reweight}
Hyperparameter $p$ controls the weight assignment in the re-weighting where a larger $p$ means a larger difference among groups. Fig.~\ref{fig:test_reweight}(a-c) shows that the impact of re-weighting on various methods is different, but they align well with the validation result. Therefore, the validation is effective in selecting a hyperparameter that matches the requirement for utility and fairness tradeoff. Generally, when $p$ increases, utility performance decreases while fairness performance increases. MF and NGCF gain a large fairness improvement with a small decrease in utility, but CAGCN$^*$ needs a larger sacrifice to obtain a larger improvement in fairness. For NGCF, we also observe an increase in fairness when enforcing a larger $p$. We hypothesize that this will also happen for the other two methods if we further increase $p$ since when utility performance becomes so low, the same quantity of performance gap would lead to larger unfairness according to the unfairness definition. Another potential reason would be that the relative order of group performance might change at some certain $p$ (i.e., previously, the majority group has better performance, and now the minority might have better performance), resulting in the enlargement of the performance gap when $p$ increases.

\subsection{Experimental Results with Re-ranking}

\subsubsection{Re-ranking Based on Baseline Models}

The dashed red line and the solid blue line in Fig.~\ref{fig:rerank_performance} correspond to the performance of baselines and re-ranking models. When $\lambda$ increases, utility and fairness performance both improve. For utility performance, MF has the largest improvement, while NGCF and CAGCN$^*$ show smaller ones. The fairness performance improves for all of them. Surprisingly, the traditional utility-fairness trade-off (i.e., fairness usually improves at the cost of utility) does not occur. We interpret this with group inconsistency analysis.

\subsubsection{Re-ranking Based on Re-weighted Models}
\begin{figure}[t]
\vspace{-1ex}
    \centering
    \includegraphics[width=0.44\textwidth]{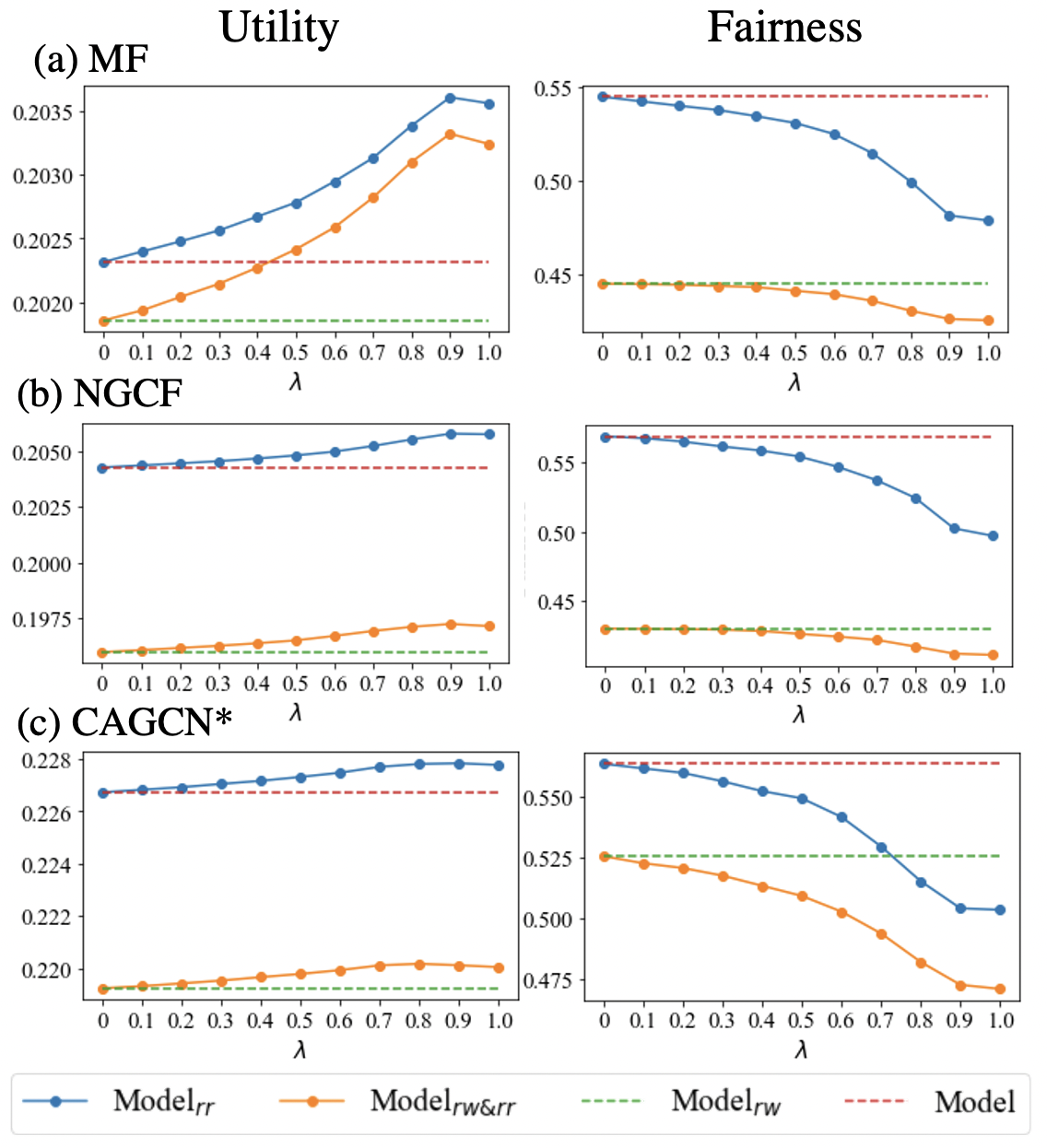}
    \caption{The utility and fairness performance of variants (1) the baseline model (Model); (2) the re-weighting model (Model$_\text{rw}$); (3) the re-ranking model (Model$_\text{rr}$); and (4) the re-ranking model based on re-weighted model (Model$_\text{rw\&rr}$).}
    \label{fig:rerank_performance}
\end{figure}

The dashed green line in Fig.~\ref{fig:rerank_performance} corresponds to the performance of the re-weighted model where the same hyperparameter is selected, and the solid orange line shows the re-ranking performance based on the re-weighted models. A similar trend is observed. Both utility and fairness improve for all the methods after re-weighting. When comparing with the same $\lambda$ without re-weighting, the utility performance of Model$_{\text{rr}}$ is lower than Model$_{\text{rw\&rr}}$ since the base re-weighted model sacrifice a little utility performance as reported. On the other hand, the re-weighted model has improved fairness, providing a good basis for re-ranking. Therefore, with the same $\lambda$, Model$_{\text{rw\&rr}}$ has better fairness than Model$_{\text{rr}}$. This result shows that the re-ranking strategy is effective irrespective of being applied to the baseline model or the re-weighted model.

\subsubsection{Interpretation from Group Calibration}
\label{sec.group_inconsistency}

\begin{figure}[t!]
    \centering
    \vspace{-1.5ex}
    \includegraphics[width=0.45\textwidth]{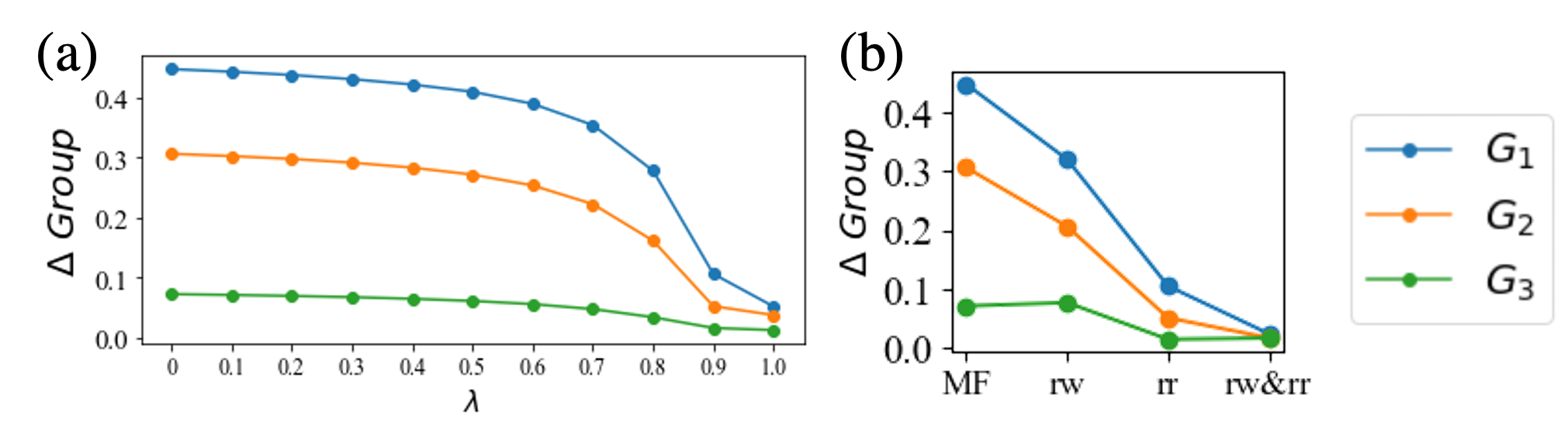}
    \caption{Analysis of group calibration on MF (a) based on different $\lambda$s; (b) based on different model variants.
    }
    \label{fig:discussion_inconsistency}
    \vspace{2ex}
\end{figure}

We take MF as a representative for analysis. Fig.~\ref{fig:discussion_inconsistency}(a) shows that the inconsistency decreases when $\lambda$ increases, affirming the re-ranking's effectiveness. The extent of improvement varies among groups. $G_3$ already has a small inconsistency before re-ranking and thus a smaller consistency improvement. Since consistency is related to recommendation quality (shown in the Appendix), $G_3$ also has a smaller performance gain. Therefore, while the overall performance increases, the gap between groups is reduced and fairness is improved, avoiding the fairness-utility tradeoff. We also explore the inconsistency of different model variants in Fig.~\ref{fig:discussion_inconsistency}(b) where the combined model has the smallest inconsistency for all groups.

\subsection{Discussion of Re-weighting and Re-ranking}
Both strategies improve fairness. Re-weighting improves fairness at a utility cost while re-ranking improves them together. We draw the following observations from the results:
\begin{itemize}[leftmargin=*]
    \item \textbf{Effect on fairness}: re-weighting outperforms re-ranking in fairness on MF and NGCF. This suggests that in-processing method, which changes the training process, may be more effective for fair recommendations. The combination Model$_\text{rw\&rr}$ achieves the best fairness.
    \item \textbf{Effect on utility}: re-weighting generally decreases utility performance, and re-ranking can improve utility performance in addition to improving fairness.
    \item \textbf{Discussion on calibration}: re-weighting, although not designed to improve calibration, reduces the inconsistency, which gives another interpretation of its effect on fairness. 
\end{itemize}
In summary, re-weighting and re-ranking strategies both have unique advantages. Re-weighting improves more on fairness, while re-ranking improves utility and better calibration. Combining them leads to even better performance. To further validate the effectiveness, we include experiments on another non-dating dataset for benchmarking purposes. Additionally, since re-ranking strategy utilizes gender feature, we conduct experiments to verify whether it is the key to success. Details of these experiments are in the Appendix.


\section{Conclusion}
\label{sec-conclusion}

Sexual orientation, which is a significant factor for individuals to find a satisfying romantic relationship, is under-investigated in online dating recommender systems. In this paper, to investigate whether users with varying preferences for the opposite gender are treated fairly by recommender systems, we leverage our proposed metric, Opposite Gender Interaction Ratio (OGIR). The empirical experiments on a real-world online dating dataset show consistent unfairness among user groups based on OGIR across algorithms, which provide better recommendations for the majority group (i.e., $G_3$ with higher OGIR) than the minority groups (i.e., $G_1$ and $G_2$ with lower OGIR). Then, based on our validated hypothesis that bias/unfairness is associated with group quantity and calibration imbalances, we propose a fair recommender system based on re-weighting and re-ranking strategies designed to alleviate the two imbalance challenges. Experimental results show that both strategies independently help improve fairness, but when combined they lead to the best overall performance in terms of maintaining utility while significantly improving fairness. 

\section{Ethics Statement}
Promoting fairness in online dating recommendations could create a more inclusive environment, where users of diverse backgrounds and preferences, such as varying sexual orientations, receive similar or equitable treatment. This not only enhances user satisfaction but also contributes to the platform's long-term sustainable development. Moreover, this work can raise awareness about the existence of bias in recommender systems and thereby encourage further fairness research in this field. Beyond the specific context of online dating, the proposed strategies of re-weighting and re-ranking (to mitigate bias associated with data and calibration imbalance issues, respectively) can be applied to other applications to promote fairness among diverse user groups.

In the studied dataset, some users do not fill in gender identity, and one potential reason besides privacy concerns could be that the platform only provides binary options and these users do not identify themselves as male/female. Valuing the importance of these users, one future direction will be looking into their characteristics and interaction patterns. We note that we advocate dating platforms to offer more gender identity options and explicitly collect information on sexual orientation to better serve users. However, we note that even if users were able to explicitly identify themselves as bisexual to the system, it is likely such unfairness across those bisexual users would still exist, especially due to the calibration imbalance, which our proposed re-ranking has been shown to help mitigate.

\section*{Acknowledgements} This research is supported by the National Science Foundation (NSF) under grant number IIS2239881. Additionally, the authors would like to thank Oldrich Neuberger for providing the data and Lukas Brozovsky and Vaclav Petricek for preparing the dataset. The authors appreciate the anonymous reviewers for dedicating their time and efforts during the review process and offering insightful and constructive feedback.

\bibliography{references}

\appendix
\section{Appendix}


\subsection{Baselines}
The baselines evaluated in this paper include seminal works and current state-of-the-art:
\begin{itemize}[leftmargin=2ex]
    \item \textbf{MF}~\cite{rendle2012bpr}: Matrix Factorization is a simple yet effective method to predict interaction scores between entities with the goal to find entity embeddings whose dot product best resembles the observed interactions.
    \item \textbf{NGCF}~\cite{wang2019neural}: Neural Graph Collaborative Filtering is the first GNN-based model to leverage the collaborative filtering (CF) effect for recommendations, which utilizes higher-order connectivities of history interactions to learn entity embeddings.
    \item \textbf{CAGCN${\bf ^*}$}~\cite{wang2022cagcn}: Collaboration-Aware Graph Convolutional Network is a fusion model of LightGCN~\cite{he2020lightgcn} and CAGCN~\cite{wang2022cagcn}. LightGCN advances NGCF by simplifying the feature propagation process, which removes the linear transformation and nonlinear activation steps. CAGCN analyzes how message-passing captures CF effect and proposes a topological metric, Common Interacted Ratio (CIR), for collaboration-aware propagation.
\end{itemize}

Although these algorithms are commonly applied to user-item interactions, they are easily leveraged for user-user interactions (e.g., friend/dating recommendations).

\subsection{Pre-processing}
To improve data quality, we perform the following: (1) filter accounts lacking self-identified gender information as $\text{OGIR}$ needs gender context; (2) filter edges with a rating less than $10$ so that the remaining edges show strong preferences; and (3) apply k-core setting iteratively to remove users with interaction number smaller than $k=5$. We randomly split the dataset into train/validation/test based on $60\%/20\%/20\%$ proportions. To further ensure that all genders have the chance to be assigned to train/validation/test, we randomly split for females and males separately.

\subsection{Discussions}
\label{sec-discussion}
In this section, we discuss the limitations and future directions. This research identifies the consistent existence of unfairness (according to the quality of personalized partner recommendations) among groups of users with varying preferences for the opposite gender across various recommendation methods, emphasizing the significance and need for fairness concerning aspects of sexual orientation on online dating platforms (especially in regard to their recommendations). Currently, the experiments are limited to one real-world dating dataset due to the scarcity of accessible online dating datasets. Furthermore, as the user profiles in the dataset did not explicitly encode sexual orientations, we have presented opposite gender interaction ratio (OGIR) to infer groups that are associated with user sexual orientations for this study. It is anticipated that with more attention drawn to this domain, additional data would be made available, enabling further verification of the proposed strategies and stimulating further advancements in the field. As an initial exploration, this research investigates fairness in online dating recommendations by focusing on three representative recommender algorithms. Other recommendation approaches (e.g., matching algorithms~\cite{hitsch2010matching}, reciprocal dating recommendations~\cite{xia2015reciprocal}, content-based recommender systems that utilize rich profile content~\cite{adomavicius2010context}) are also worth investigating and we encourage such follow-up work from the community. The presented work concentrates on one sensitive feature (OGIR, which is associated with user sexual orientation), but designing a fair model that takes into account the fairness of multiple sensitive features is also crucial, which we leave as future work.

\subsection{Group Calibration}
\begin{figure}[t!]
    \centering
    \includegraphics[width=\columnwidth]{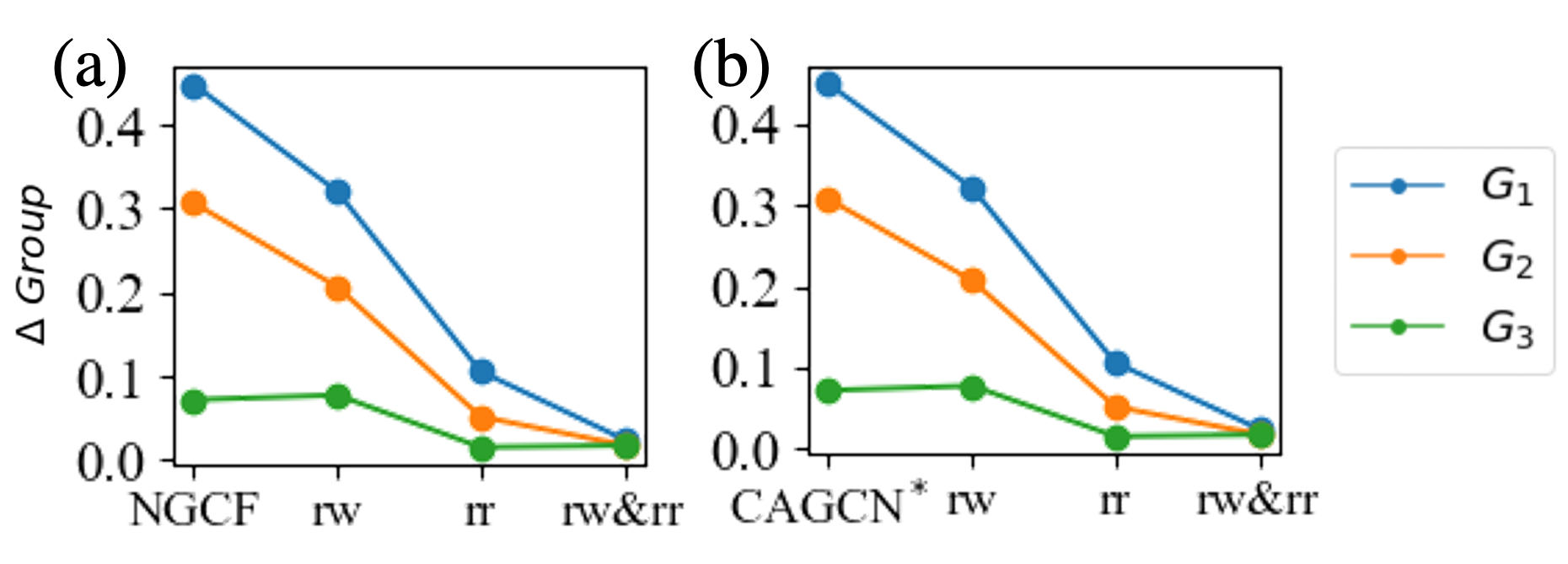}
    \vspace{-2.5ex}
    \caption{Group calibration on NGCF and $\text{CAGCN}^*$.
    }
    \label{fig:calibration}
    \vspace{2.5ex}
\end{figure}

In this section, we show the group calibration results for NGCF and $\text{CAGCN}^*$ in Fig.~\ref{fig:calibration} and they show the same trend as for MF. The re-ranking model based on the re-weighted baseline achieves the smallest inconsistency. Additionally, the re-weighting model generally achieves better consistency when compared with the base model. This indicates that although re-weighting is not designed for better calibration, it will contribute to better consistency between the historical interaction and recommendation list.

\subsection{Time Complexity Analysis}
The re-weighting strategy will not alter the time complexity of the baseline model after its appliance. This is due to the fact that the additional weight computation will be only executed once and its complexity is negligible during the analysis. For re-ranking strategy, it will be applied to each user. When greedily picking the next user to be recommended, a comparison of the score between the female and male user with the highest relevance will be executed which takes constant time. Since the recommendation problem itself requires candidate list sorting to obtain top $K$, the extra time for re-ranking with the sorted list for each user is $O(|\mathcal{C}|)$, i.e., iterating over the candidates (although most scenarios expected closer to $K$), which does not significantly impact the complexity. Therefore, both strategies are time-efficient.

\subsection{Hyperparameters}
For all methods, we train with $0.001$ learning rate and $64$ embedding dimension. For $\text{CAGCN}^*$, we use `jc' as the metric and set coefficient to 1. The training epoch is set to 200. The hyperparameter $p$ of re-weighting strategy, which controls the power in the weight denominator, is tuned from [0, 0.5, ..., 2.5]. The best hyperparameter is chosen based on the validation curve and the final hyperparameters are 1.5 for MF, 1.0 for NGCF, and 0.5 for $\text{CAGCN}^*$. The hyperparameter $\lambda$ for re-ranking strategy, which controls the tradeoff between utility and fairness consideration, is tuned from [0, 0.1, ..., 1.0]. We report all the results without choosing a specific hyperparameter.

\subsection{Incorporating Features}
\label{sec.incorprating_features}
Due to privacy concerns, the user profiles such as the photographs or their descriptions are unavailable in the studied dataset - only gender information is available. While current RS indeed leverages rich user features, no modern dating platforms were willing to share their data for privacy concerns, but we encourage features to be included in practice.

\begin{figure}
    \centering
    \includegraphics[width=\columnwidth]{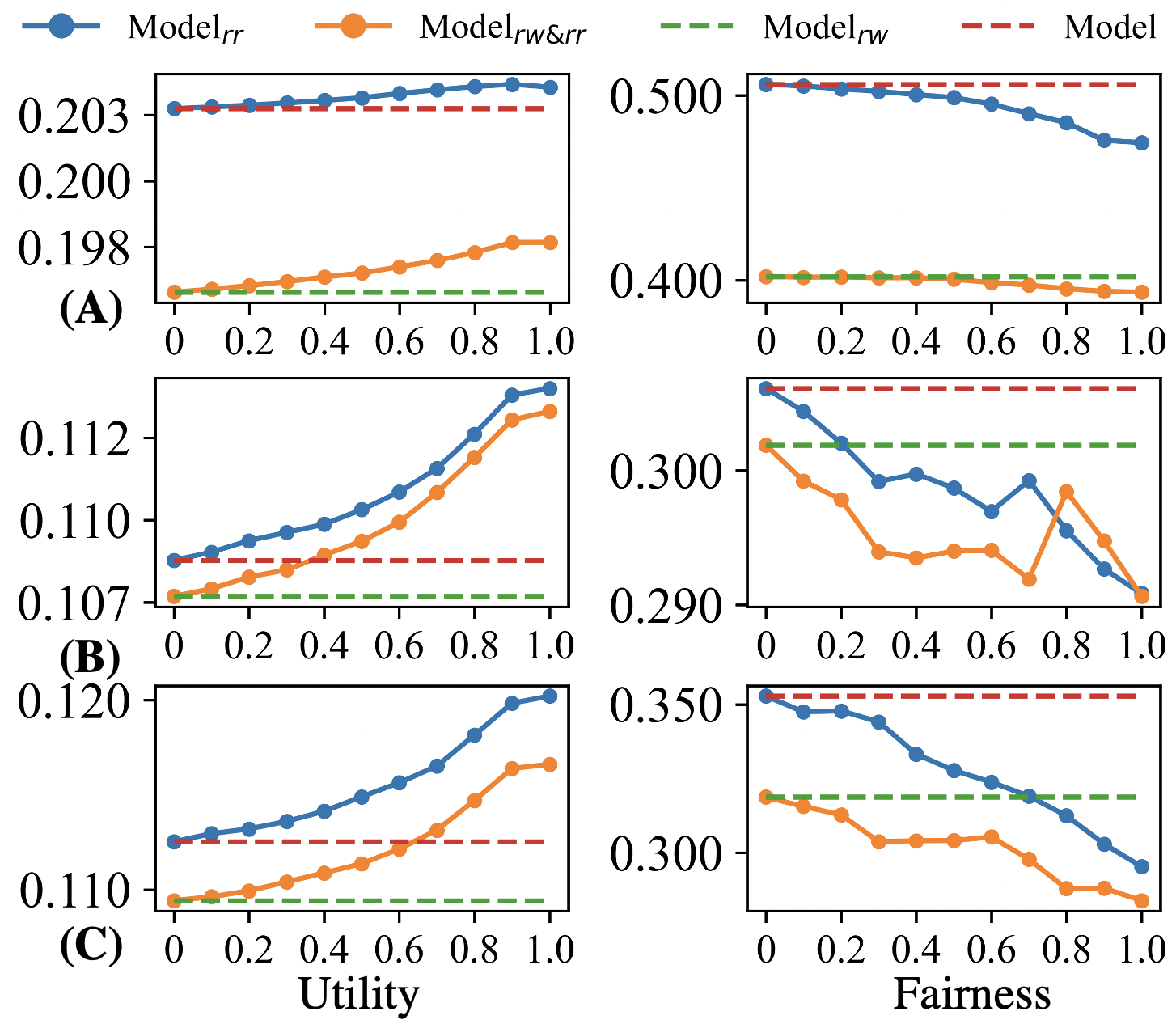}
    \caption{Utility-Fairness trade-off (A) MF training with gender information; (B)/(C) MF/NGCF results for a pseudo dating network dataset, \textit{deezer-europe} (a social network with gender information) for benchmarking purposes.}
    \vspace{2.5ex}
    \label{fig:rebuttal}
\end{figure}

To investigate whether incorporating gender information is the key to the success of re-ranking, we conducted an experiment where the last two dimensions of initial embeddings are assigned to the one-hot gender encoding (i.e., to inject gender information). The results on MF in Fig.~\ref{fig:rebuttal}(A) show that: (1) added gender features provide no significant utility increase; (2) re-weight and re-ranking in this setting have similar observations as those without gender features.

\begin{figure*}[t]
    \centering
    \includegraphics[width=\textwidth]{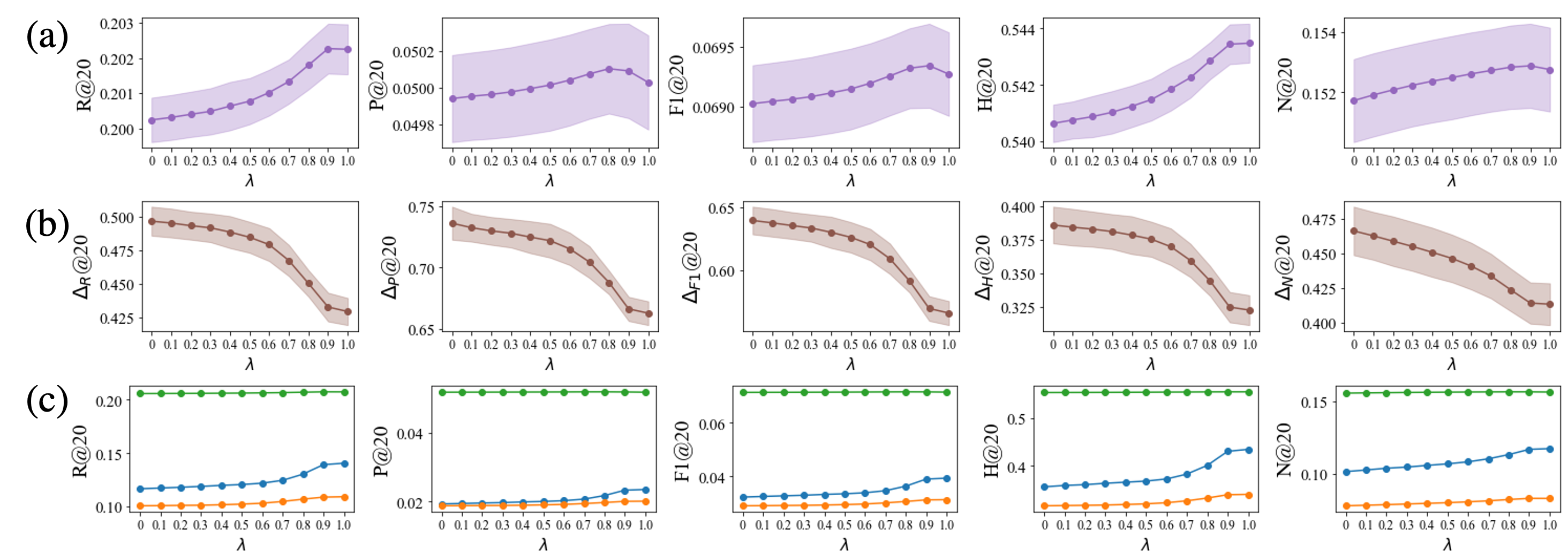}
    \caption{Re-ranking results of different $\lambda$s on MF (a) utility performance; (b) fairness performance; (c) group utility performance where blue/orange/green relate to groups $G_1$/$G_2$/$G_3$. Groups are divided based on their opposite gender interaction ratio (i.e., $G_1=\{u|\text{OGIR}_u\in [0, \frac{1}{3})\}$ with $G2$, $G3$ similarly defined.}
    \label{fig:MF}
    \vspace{1ex}
\end{figure*}

\begin{figure*}[t]
    \centering
    \includegraphics[width=\textwidth]{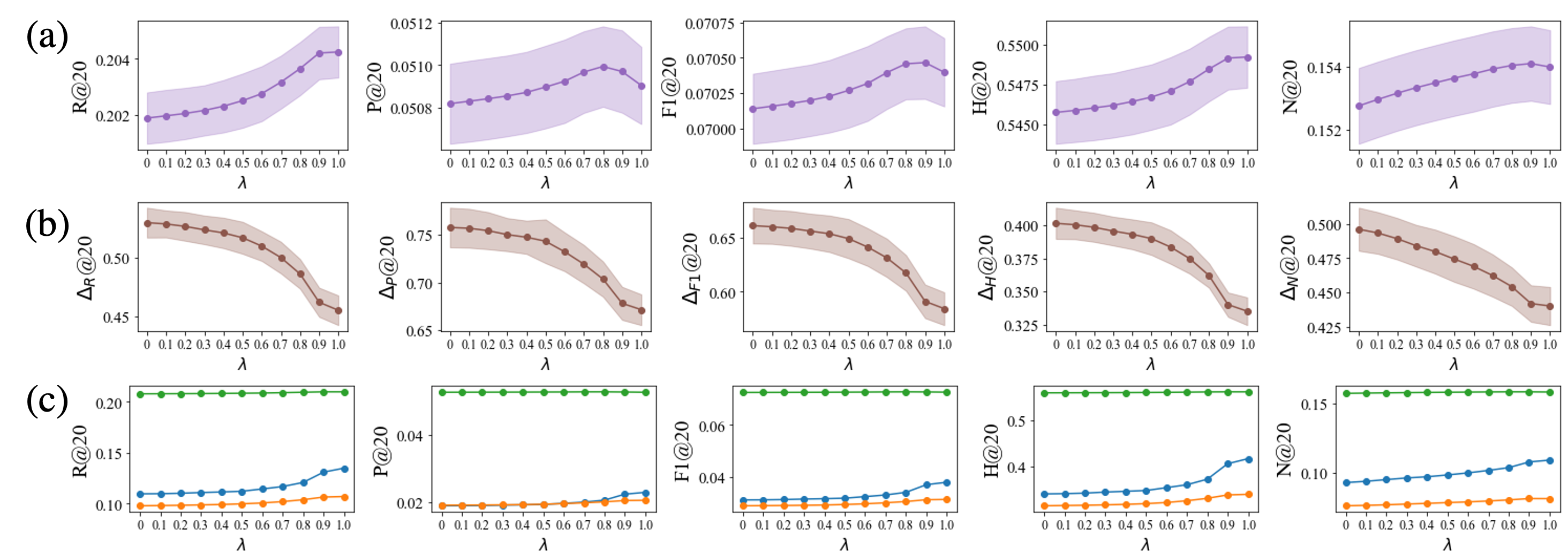}
    \caption{Re-ranking results of different $\lambda$s on NGCF.}
    \label{fig:NGCF}
    \vspace{1ex}
\end{figure*}

\subsection{Other Datasets}
\label{sec.other_datasets}
Due to privacy concerns, only one dating dataset that contains interactions and gender is available. To further validate our methodology, we seek other datasets for benchmarking purposes in addition to the online dating application.

We found such a dataset called \textit{deezer-europe}~\cite{rozemberczki2020characteristic} which also has gender information. It is a social network of users on Deezer from European countries, where edges represent mutual follower relationships. We use this dataset to test the effectiveness for benchmarking purposes. Note that the application setting here does not have the ethical meaning as in the dating dataset. Specifically, given the dataset, we calculated the OGIR based on the interaction the same way as reported in the paper. To mimic the data statistic in the dating dataset, we use fixed OGIR thresholds to divide the groups so that an imbalanced group distribution similar to the dating scenario is obtained. Results in Fig.~\ref{fig:rebuttal}(B)(C) on \textit{deezer-europe} align with the findings in the paper. This further validates the effectiveness of our proposed strategies.

\pagebreak 
\subsection{Sensitivity Analysis of $\lambda$}
\begin{figure*}[t]
    \centering
    \includegraphics[width=\textwidth]{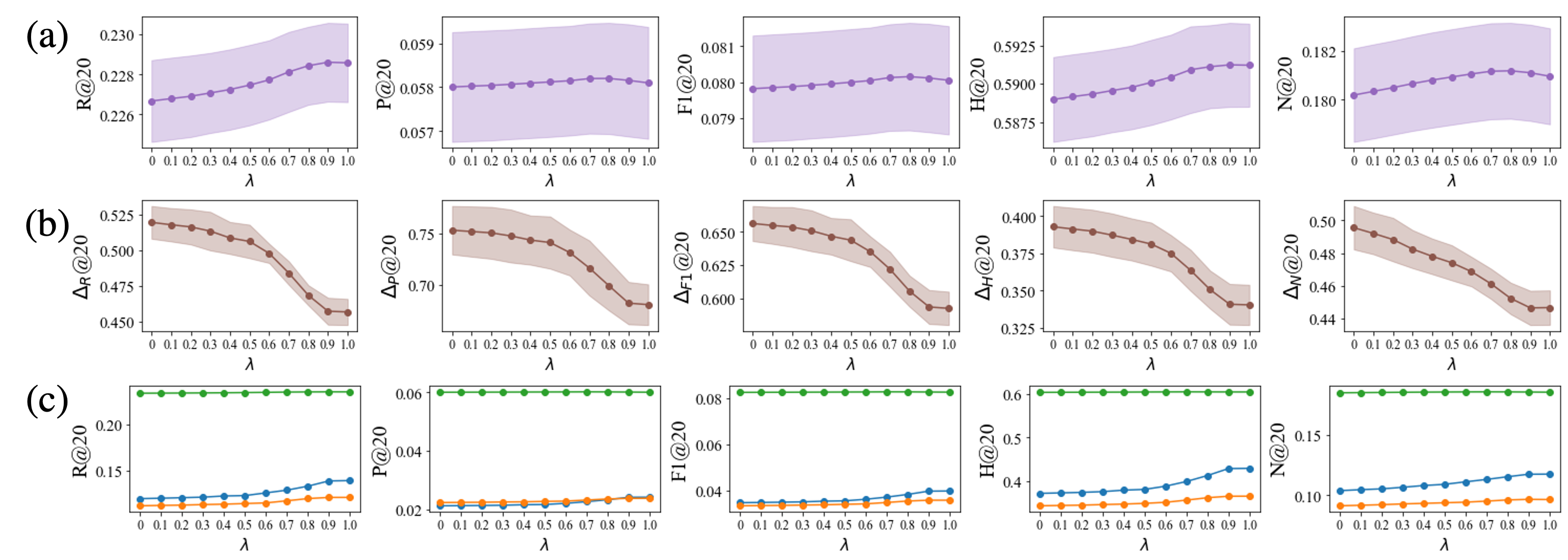}
    \caption{Re-ranking results of different $\lambda$s on $\text{CAGCN}^*$.}
    \label{fig:CAGCN}
\end{figure*}

In this section, we conduct a comprehensive analysis of the hyperparameter $\lambda$. The results for different models are shown in Fig.~\ref{fig:MF} to Fig.~\ref{fig:CAGCN}. Take MF as an example for analysis, Fig.~\ref{fig:MF}(a)-(b) suggest that generally, both utility and fairness improve along with the increase of $\lambda$ in every metric. A closer look at the performance per group in Fig.~\ref{fig:MF}(c) shows that the performance change for $G_3$ is much more stable than $G_1$ and $G_2$, as $G_1$ and $G_2$ have a greater improvement when $\lambda$ increases. This is related to the different inconsistency effects as shown in Fig.~\ref{fig:discussion_inconsistency}. $G_3$ has a small improvement in the inconsistency and therefore a small improvement in the performance.

\end{document}